\documentclass[iop,apj]{emulateapj}
\usepackage{amsmath,amssymb}
\usepackage{graphicx}
\usepackage{epstopdf}
\usepackage{natbib}
\usepackage{color}
\usepackage[breaklinks,colorlinks=true,urlcolor=blue,linkcolor=red,citecolor=blue]{hyperref}
\graphicspath{{./}{figs/}}

\newcommand{\angstrom}{{\rm \mathring A}}

\begin{document}
\title{
More variable quasars have stronger emission lines
}
\author{Wen-Yong Kang\altaffilmark{1,2}, Jun-Xian Wang\altaffilmark{1,2}, Zhen-Yi Cai\altaffilmark{1,2}, Wen-Ke Ren\altaffilmark{1,2}}
\affil{
$^1$CAS Key Laboratory for Researches in Galaxies and Cosmology, University of Science and Technology of China, Chinese Academy of Sciences, Hefei, Anhui 230026, China; kwy0719@mail.ustc.edu.cn, jxw@ustc.edu.cn\\
$^2$School of Astronomy and Space Science, University of Science and Technology of China, Hefei 230026, China
}

\begin{abstract}
The UV/optical variation, likely driven by accretion disc turbulence, is a defining characteristic of type 1 active galactic nuclei (AGNs) and quasars.
In this work we investigate an interesting consequence of such turbulence using quasars in SDSS Stripe 82 for which the measurements of the UV/optical variability amplitude are available from $\sim$ 10 years long light curves.
We discover positive correlations between  UV/optical variability amplitude $\sigma_{rms}$ and equivalent widths of CIV, Mg II and [OIII]5007 emission lines.
Such correlations remain statistically robust through partial correlation analyses, i.e., after controlling the effects of other variables including bolometric luminosity, central supermassive black hole mass, Eddington ratio and redshift. This, for the first time, indicates a causal link between disc turbulence and emission line production. 
We propose two potential underlying mechanisms  both of which may be involved: 1) quasars with stronger disc turbulence have on average bluer/harder broadband SED, an expected effect of the disc thermal fluctuation model; 2)
stronger disc turbulence could lead to launch of emission line regions with larger covering factors. 

\keywords{accretion, accretion disks -- galaxies: active -- quasars: general -- quasars: emission lines}
\end{abstract}

\section{Introduction}
\label{introduction}
The presence of prominent optical/UV broad emission lines (BELs) is a defining characteristic of type 1 active galactic nuclei (AGNs) and quasars. 
It is widely accepted that the BEL emitting clouds in the broad line region (BLR) are photoionized by the central radiation, thus BELs are important probes of the ionizing continuum and subsequently the inner accretion disc where ionizing photons are produced. 
Meanwhile, the BLR clouds themselves may physically originate from the disc in forms of winds or failed winds driven by various potential mechanisms \citep[e.g.][]{1992ApJ...385..460E,1994ApJ...434..446K,1995ApJ...451..498M,2004ApJ...616..688P,2011A&A...525L...8C,2018MNRAS.474.1970B},  though non-disc origin models also exist \citep[e.g.][]{2017NatAs...1..775W}. 

An intimately linked phenomenon is the well-known anti-correlation between the BEL equivalent width ($EW$) and continuum luminosity, the so called ``Baldwin effect" \citep{1977ApJ...214..679B}. The Baldwin effect of various BELs\footnote{There also exists Baldwin effect for narrow emission lines \citep[e.g.][]{1992ApJS...80..109B,2007ASPC..373..355S, 2014Natur.513..210S}, as well X-ray Fe K$\alpha$ line \citep[e.g.][]{1993ApJ...413L..15I, Jiang2006, Shu2012, 2013A&A...553A..29R}.} has been extensively investigated for over four decades \citep[e.g.][]{1984ApJ...276..403W, 1989ApJ...338..630B, 1992MNRAS.254...15N, 1995MNRAS.274..504F, 2002ApJ...581..912D, 2008MNRAS.389.1703X, 2009ApJ...703L...1D, 2009ApJ...702..767W, 2010ApJS..189...15K, 2012MNRAS.427.2881B, 2015ApJ...805..124S}, 
however the physical origin of the anti-correlations and the notably large scatter in the line $EW$ are still under debates. This is likely because the observed line $EW$ could be influenced by many factors,  including the broadband spectral energy distribution, metallicity, BLR covering factor and ionization, etc.  Besides, the disc inclination effect (the limb darkening and projected disc surface area effects, e.g., \citealt{Risaliti2011,Zhang2013}) and the continuum variation \citep[e.g.][]{Jiang2006,Shu2012} may produce artificial anti-correlations between line $EW$ and continuum luminosity in AGN samples.
Searching for other such factor(s) may yield new clues to understanding the BEL production.

Aperiodic multi-band flux variation is another notable characteristic of AGNs \citep[e.g.][]{1997ARA&A..35..445U}.  
In UV/optical, the variation is generally attributed to thermal fluctuations in the accretion disc, likely driven by magnetic turbulence \citep{Kelly2009}, a theoretically critical process but observationally hard to probe.
Besides studying the correlations with physical parameters including luminosity, wavelength, Eddington ratio, black hole mass and redshift \citep{berk2004ensemble, Wilhite2005, Wold2007, wilhite2008variability, Bauer2009, 2010ApJ...716L..31A, Macleod2010, 2011A&A...525A..37M, Zuo2012, 2013A&A...560A.104M, Kozlowski2016ApJ826}, exploring additional parameters correlating with variability could help to reveal the consequences of the magnetic turbulences \citep[e.g. X-ray loudness in][]{2018ApJ...868...58K}.
 
 It is intriguing to note that, similar to BEL $EW$, the UV/optical variability amplitude in AGNs also anti-correlate with luminosity \citep[e.g.][]{berk2004ensemble,2013A&A...560A.104M,2018ApJ...868...58K}.
 Is there any intrinsic and physical link between the two fundamental characteristics of AGNs?
Considering both BEL production and UV/optical variability are closely related to processes in the accretion disc, observationally revealing such a link would be useful to probe the yet-vague underlying mechanisms.

In this work we present a first exploratory study of the potential intrinsic correlation between the BELs (as well as [OIII]5007) and UV/optical variability. We focus on the intrinsic correlation between line $EW$s and UV/optical variability amplitudes, which could be precisely measured for a large sample of quasars.
 In \S\ref{S:sample} we present the quasar sample and the quantities utilized in this study. We perform partial correlation analyses in \S\ref{S:correlation} to reveal the intrinsic correlations between the $EW$ (of various lines) and UV/optical variability amplitude, controlling the effects of other variables including Eddington ratio, supermassive black hole mass and redshift.  In \S\ref{S:discussion} we propose and discuss two potential mechanisms for the intrinsic link we discovered. Throughout this work, cosmological parameters of $H_0=70 km\cdot{}s^{-1}\cdot{}Mpc^{-1}$, $\Omega_m=0.3$ and $\Omega_{\Lambda}=0.7$ are adopted. 
 
\section{The Quasar Samples}
\label{S:sample}

SDSS Stripe 82, which has been scanned over 60 times in five bands ($ugriz$) by the Sloan Digital Sky Survey, is a 290 $deg^2$ equatorial field of the sky \citep{Sesar2007}. Recalibrated 10-year-long SDSS light curves in $ugriz$ for 9275 spectroscopically confirmed quasars in Stripe 82 were presented by \cite{Macleod2012}. Their physical parameters, including bolometric luminosity, black hole mass, redshift, and emission line properties ($FWHM$, flux, $EW$) could be extracted from \cite{shen2011catalog}.
Such a large sample of quasars is adopted in this work to explore the intrinsic relation between emission lines and UV/optical variation.
 
We note that many studies adopted the damped random walk process to model quasar light curves \citep{Kelly2009, Macleod2010, Zu2013, Kozlowski2010} with two parameters, $\tau$ (the characteristic timescale) and $SF_{\infty}$ (the structure function). However, due to the limited length and the sparse sampling, for many quasars these parameters are poorly constrained with SDSS Stripe 82 light curves \citep{Kozlowski2017}. 
In this work, similar to \cite{2018ApJ...868...58K}, we quantify the intrinsic variability amplitude of each source in each band with a single model-independent parameter, i.e., the excess variance $\sigma_{rms}$ \citep{2003MNRAS.345.1271V}
 \begin{align}
  \sigma_{rms}^2=\frac{1}{N-1}\sum(X_i-\bar{X})^2 - \frac{1}{N}\sum\sigma_i^2
   \label{e2_1}
 \end{align}
 where $X_i$ is observed magnitude, $\bar{X}$ the average magnitude, $\sigma_i$ the photometric uncertainty of each observation, and $N$ the number of photometric measurements. If there is no intrinsic variation, the expected value of $\sigma_{rms}$ is zero with a statistical uncertainty \citep{2003MNRAS.345.1271V} of 
 \begin{align}
  err(\sigma_{rms}^2)=\sqrt{\frac{2}{N}}\times\frac{1}{N}\sum\sigma_i^2
  \label{e2_2}
 \end{align}
 We dropped $u$ and $z$ band in which the photometric uncertainties are significantly larger comparing with the other three bands. 
 
 In this work we focus on the most prominent lines in SDSS spectra, including broad MgII, CIV, broad H$\beta$, as well as the narrow emission line [OIII]5007. 
 We build samples for each line within a certain redshift range.
  
 For MgII line, we select quasars from \cite{shen2011catalog} with broad MgII measurements\footnote{Two sources with un-physically large values of MgII $EW$ ($>$ 20,000$\angstrom$) are excluded.} (0.35 $<$ z $<$ 2.25) and median SDSS spectral $S/N$ per pixel $\ge$ 3 in the restframe $2700-2900\angstrom$.
 The MgII sample includes 6553 quasars, for which we adopt the Virial black hole mass based on MgII (S10 in \citealt{shen2011catalog}) and bolometric luminosity derived from $L_{3000\angstrom}$. 
 
 The CIV sample\footnote{Note \cite{shen2011catalog} did not subtract a narrow component while fitting the CIV line.} contains 3313 quasars (1.50 $<$ z $<$ 3.69) with median SDSS spectral $S/N$ per pixel $\ge$ 3 in the restframe $1500-1600\angstrom$.
 For this sample, the CIV derived Virial black hole mass as VP06 from \cite{shen2011catalog} and bolometric luminosity based on $L_{1350\angstrom}$ are adopted. Note that CIV line based BH mass could be significantly biased \citep[e.g.][]{Coatman2016, Coatman2017}.
 
 Both broad H$\beta$ and [OIII]5007 samples are required to have median SDSS spectral $S/N$ per pixel $\ge$ 3 in the restframe $4750-4950\angstrom$, 
 including 1226 (0.08 $<$ z $<$ 0.89)\footnote{Four sources are dropped because of unreasonably large H$\beta$ $EW$ (above 1000$\angstrom$).} and 1132 (0.08 $<$ z $<$ 0.84) quasars respectively.
For both samples, which indeed largely overlap, the broad H$\beta$ based Virial black hole mass (VP06 in \citealt{shen2011catalog}) and $L_{5100\angstrom}$-based bolometric luminosity are adopted.

\label{S:31}
\begin{figure}%[!h]
  \centering
  \includegraphics[width=\linewidth]{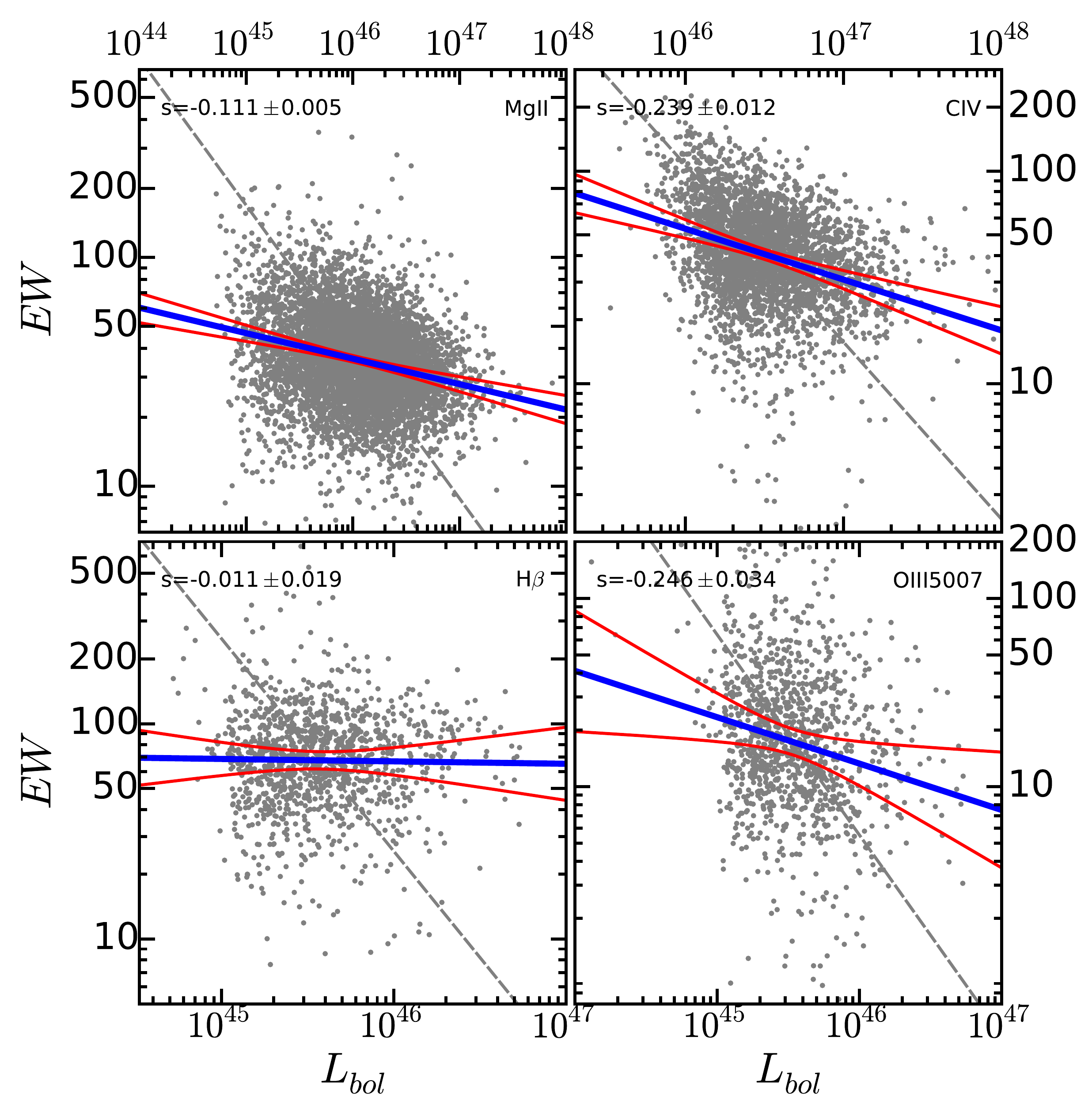}
  \caption{Correlations between line $EW$ and bolometric luminosity for the four samples. 
 Blue and red lines plot the best-fit standard linear regression and the corresponding 6$\sigma$ confidence bands. 
The best-fit linear regression slopes with 1$\sigma$ uncertainties, $s$, are given in the upper left corner in each panel, and the line names  in the upper right corner. The $EW$ of Mg II, CIV and OIII (but not H$\beta$) clearly anti-correlates with $L_{bol}$.
Note the standard linear regression slopes are obtained taking x-axis as the independent variable (hereafter the same). For reference, grey dashed lines plot the bisector regression results.  
}
  \label{Fig3_1}
\end{figure}

\begin{table*}[h]
  \centering
  \caption{Correlations coefficients and linear regression slopes between line $EW$ and other parameters (named in the left-most column).}
  \begin{tabular}{c c c c c c}
    \hline
    \hline
       &&broad MgII&CIV&broad H$\beta$&[OIII]5007\\
    \hline
       \multicolumn{6}{ c }{Pearson's Rank apparent correlation coefficients $r$, confidence levels $rcc$ and linear regression slopes $s$ }\\
    \hline
       &r&-0.259(0.014)&-0.325(0.007)&-0.017(0.018)&-0.212(0.077)\\
       $L_{bol}$&rcc&$<$1e-16($<$1e-16,$<$1e-16)&$<$1e-16($<$1e-16,$<$1e-16)&0.28(0.11,0.49)&3e-13($<$1e-16,2e-6)\\
       &s&-0.111$\pm$0.005&-0.239$\pm$0.012&-0.011$\pm$0.019&-0.246$\pm$0.034\\
    \hline
       &r&-0.416(0.018)&-0.118(0.014)&-0.215(0.024)&-0.196(0.039)\\
       $\frac{L_{bol}}{L_{Edd}}$&rcc&$<$1e-16($<$1e-16,$<$1e-16)&4e-12(1e-14,1e-9)&1e-14($<$1e-16,8e-12)&2e-11(5e-16,5e-8)\\
       &s&-0.209$\pm$0.006&-0.070$\pm$0.010&-0.106$\pm$0.014&-0.159$\pm$0.024\\
    \hline
       &r&0.094(0.013)&-0.132(0.012)&0.197(0.029)&0.046(0.042)\\
       $M_{bh}$&rcc&1e-14($<$1e-16,2e-11)&1e-14($<$1e-16,2e-12)&2e-12(6e-16,2e-9)&0.06(2e-3,0.45)\\
       &s&0.039$\pm$0.005&-0.072$\pm$0.009&0.094$\pm$0.013&0.037$\pm$0.024\\
    \hline
       &r&-0.044(0.012)&-0.057(0.007)&0.098(0.047)&-0.084(0.044)\\
       $1+z$&rcc&2e-4(3e-6,5e-3)&6e-4(1e-4,2e-3)&3e-4(2e-7,0.04)&2e-3(8e-6,0.09)\\
       &s&-0.096$\pm$0.027&-0.260$\pm$0.080&0.537$\pm$0.155&-0.813$\pm$0.287\\
    \hline
       \multicolumn{6}{ c }{Partial correlation coefficients $r$ and confidence levels $rcc$}\\    
    \hline
       $\frac{L_{bol}}{L_{Edd}}(M_{bh},1+z)$&r&-0.454(0.020)&-0.314(0.018)&-0.141(0.030)&-0.226(0.044)\\
       &rcc&$<$1e-16($<$1e-16,$<$1e-16)&$<$1e-16($<$1e-16,$<$1e-16)&4e-7(9e-10,5e-5)&7e-15($<$1e-16,4e-10)\\
    \hline
       $M_{bh}(\frac{L_{bol}}{L_{Edd}},1+z)$&r&-0.239(0.017)&-0.317(0.017)&-0.007(0.028)&-0.140(0.041)\\
       &rcc&$<$1e-16($<$1e-16,$<$1e-16)&$<$1e-16($<$1e-16,$<$1e-16)&0.40(0.11,0.23)&1e-6(4e-10,4e-4)\\
    \hline
       $1+z(\frac{L_{bol}}{L_{Edd}},M_{bh})$&r&0.218(0.014)&0.098(0.009)&0.101(0.046)&-0.001(0.032)\\
       &rcc&$<$1e-16($<$1e-16,$<$1e-16)&9e-9(4e-10,2e-7)&2e-4(1e-7,0.03)&0.48(0.13,0.15)\\
    \hline
       \multicolumn{6}{ c }{Multiple linear regression slopes (see Equation \ref{e:EW})}\\    
    \hline
       $\frac{L_{bol}}{L_{Edd}}$&a&-0.327$\pm$0.008&-0.283$\pm$0.015&-0.114$\pm$0.023&-0.310$\pm$0.040\\
       $M_{bh}$&b&-0.152$\pm$0.008&-0.266$\pm$0.014&-0.006$\pm$0.024&-0.193$\pm$0.041\\
       $1+z$&c&0.666$\pm$0.037&0.468$\pm$0.083&0.647$\pm$0.182&-0.014$\pm$0.321\\
    \hline
  \end{tabular}  
  \\
  \raggedright
%  This table lists two statistical analyses of MgII, CIV, H$\beta$ and [OIII]5007 line detected sample, respectively. 
Here $r$ and $rcc$ represent correlation coefficient and confidence level of the correlation. In the left-most column, single parameter name stands for apparent Pearson's Rank correlation of $EW_{line}$ and this parameter, while X(Y, Z) denotes partial correlation between $EW_{line}$ and X by controlling Y and Z. Values in parentheses after $r$ and $rcc$ give the standard deviations to $r$ derived from Monte Carlo simulations (see text for details), and the 1$\sigma$ confidence range of $rcc$, respectively.  %and significance level after rcc when the 1$\sigma$ uncertainty is added to and subtracted from the correlation coefficient, respectively. 
$s$ represents the best-fit apparent linear regression slope  between $EW_{line}$ and other parameters, while $a$, $b$ and $c$ are slopes of the best-fit multiple linear regression in Equation \ref{e:EW}.
  \label{Tab3_1}
\end{table*}

\section{Correlation Analyses}
\label{S:correlation}

\subsection{The Baldwin effect}

Before we look into the correlations between emission lines and UV/optical variations, we first examine the Baldwin effect in our quasar samples, an issue closely relevant to this study.
In Table \ref{Tab3_1}, we present the Pearson's Rank correlations coefficients ($r$, $rcc$) and linear regression slopes ($s$) between
line $EW$s and parameters including bolometric luminosity, Eddington ratio $L_{bol}$/$L_{Edd}$, black hole mass $M_{bh}$, and $1+z$. 
We perform Monte Carlo simulations \citep[e.g.][]{2015ascl.soft04008C, 2019MNRAS.tmp.3066T} to quantify the statistical errors of the correlation coefficient $r$ and $rcc$ due to uncertainties in the data.
This was done through adding randomized Gaussian errors to the observed parameters of each source, and performing the Pearson's Rank correlation analyses on the simulated data set. We repeat this process 100 times and calculate the standard deviation of the derived coefficients. 
In Fig. \ref{Fig3_1}, we plot $EW$ vs $L_{bol}$ result of four samples, with the slopes of the best-fit linear regression given in the upper left corner of each panel. 
 Note in this work, when performing linear regression, we adopt the standard approach simply using x-axis as the independent variable. 
This is because 1) the best-fit standard linear regression slope is directly comparable with those derived from multiple linear regression (before versus after correcting the effects of other parameters); 2) the slope is directly comparable with literature studies which adopted the standard approach, particularly those measured line $EW$ from the composite spectra at different luminosity bins \citep[e.g.][]{2002ApJ...581..912D}.
The best-fit bisector regression is plotted in figures for reference. 

 Since the luminosity, SMBH mass and redshift in our samples are clearly degenerate, i.e., quasars at high redshifts tend to be more luminous and thus have more massive black holes, 
partial correlation analyses are required to reveal the intrinsic correlation between line $EW$ and each physical parameter by controlling the effects of the others. 
We further perform partial correlation analyses
between $EW$ and each of the three parameters (Eddington ratio $L_{bol}$/$L_{Edd}$, black hole mass $M_{bh}$, and $1+z$) by controlling the other two (see Table \ref{Tab3_1}). 
Note since $L_{bol}$ is simply the arithmetic product of the Eddington ratio and the black hole mass, we need to drop it
during partial correlation analyses.  
We also adopt multiple linear regression analysis to quantify the relations between $EW$ and these three physical parameters:
 \begin{align}
 %\label{e3}
EW_{line} \sim (L_{bol}/L_{Edd})^a{M_{bh}}^b(1+z)^c
 \label{e:EW}
 \end{align}
The best-fit slopes, showing correlation patterns between those parameters consistent with the results from partial correlation analyses,  are also presented in Table \ref{Tab3_1}.

Our samples show significant Baldwin effects (the anti-correlation between line $EW$ and bolometric luminosity) in broad MgII, CIV and [OIII]5007, but no such effect in Balmer line (broad H$\beta$), consistent with literature studies \citep[e.g.][]{1984ApJ...276..403W, 1989ApJ...338..630B, 1992MNRAS.254...15N, 1999AJ....118.2658S, 2002ApJ...581..912D, 2007ApJ...671..104L, 2008MNRAS.389.1703X, 2009ApJ...703L...1D, 2009ApJ...702..767W, 2012MNRAS.427.2881B, 2017A&A...603A..49R}. 

Negative correlations between $EW$ and $L_{bol}$/$L_{Edd}$ are significant in all four emission lines, and remain evident after controlling the effects of $M_{bh}$ and $1+z$ (see Table \ref{Tab3_1}). This reveals that Eddington ratio has an intrinsic and dominant effect on $EW$, 
consistent with previous studies 
\citep[e.g.][]{2004MNRAS.350L..31B, 2005MNRAS.356.1029B, 2008MNRAS.389.1703X, 2009ApJ...703L...1D, 2012MNRAS.427.2881B}. 
Partial correlation analyses also reveal clear intrinsic anti-correlation between $EW$ and $M_{bh}$ for all lines but H$\beta$, showing $M_{bh}$ also plays a non-negligible role. 

\begin{figure}%[!h]
  \centering
  \includegraphics[width=\linewidth]{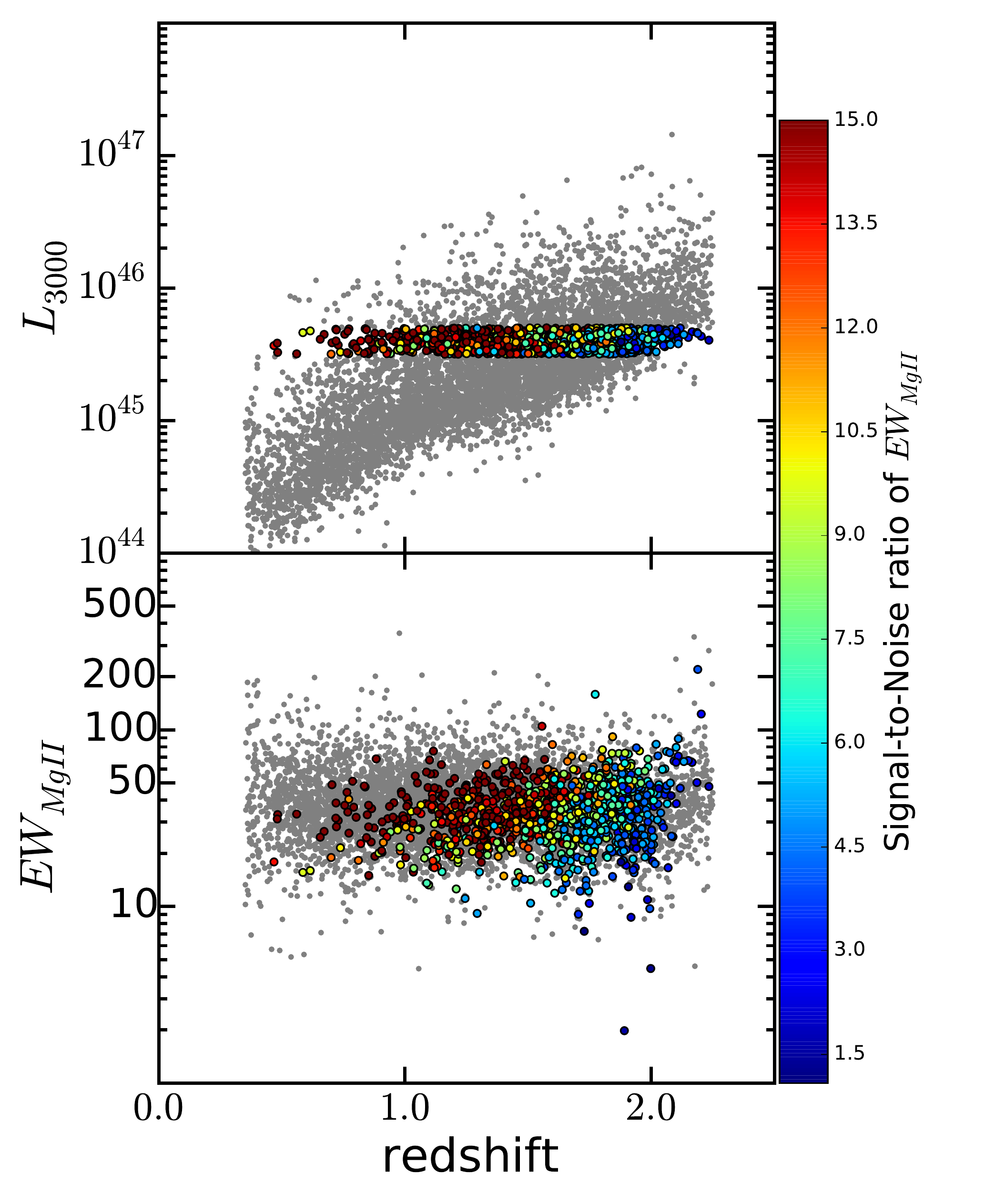}
  \caption{$L_{3000}$  (upper panel) and $EW_{MgII}$ (lower) versus redshift for the MgII sample. The quasars within a small range of $L_{3000}$ ($10^{45.5}$ to $10^{45.7}$ $erg$ $s^{-1}$) are color coded according to their broad MgII line $S/N$, demonstrating that the sample could be incomplete for low line $EW$ sources at higher redshifts.}
  \label{Fig4_1}
\end{figure}

Meanwhile, while we see no strong apparent anti-correlations between line $EW$ and $z$, consistent with \cite{2002ApJ...581..912D}, 
significant positive partial correlation between line $EW$ and $1+z$ are visible for all lines but [OIII]5007. 
This could primarily be due to a hidden selection bias of the quasar samples.  SDSS quasars were primarily color selected and spectroscopically identified based on detection of broad emission lines.
At given bolometric luminosity and SMBH mass (which means at given continuum luminosity and broad line width), quasars at higher redshifts have lower signal to noise ratio ($S/N$) in their SDSS spectra thus sources with smaller broad emission line $EW$s may have not been spectroscopically identified. 
Such selection effect actually had been noticed for a long time, and could strengthen the observed Baldwin effect for optically selected incomplete samples, since quasars with lower luminosities and lower line $EW$ are more likely to be missed from such samples \citep[e.g.][]{Osmer1980,Steidel1991}.
To demonstrate this effect in our sample, we plot the MgII sample in Fig. \ref{Fig4_1} for instance. 
Within a narrow range of continuum luminosity we can clearly see that quasars at higher redshifts tend to have smaller broad MgII line $S/N$, therefore the quasar sample could be significantly incomplete for low line $EW$ quasars at higher redshifts, yielding artificial partial correlation between line $EW$ and redshift.  
This scenario is also supported by the non-detection of the partial correlation between [OIII]5007 $EW$ and redshift, as spectroscopical identification of quasars does not rely on significant detection of the narrow line [OIII]5007.  
 An extensive study of the potential correlation between line $EW$\footnote{Such study should not be limited to the quasars in SDSS Stripe 82.} and redshift is beyond the scope of this work. 
 We stress that the aim of this work is to explore the intrinsic correlation between line $EW$ and $\sigma_{rms}$ (see \S\ref{S:33}) through partial correlation analyses, i.e., removing the effects of other parameters including redshift. Therefore,  such observational bias or the intrinsic correlation between line $EW$ and redshift (if there is any) would not affect the main results of this work, as the effect of redshift has been excluded during the partial correlation analyses below. 

 Similarly, the sample completeness could be line width dependent, as detecting a broader line requires higher $S/N$ or line $EW$, compared with a narrower line. Since the SMBH mass is derived from line width and luminosity, this effect may bias the correlation between line $EW$ and SMBH mass (or Eddington ratio). Again, such effect would not affect the main results of this work, as during the partial correlation the effects of $L_{bol}$, $M_{bh}$, Eddington ratio (thus also line width) has been removed. 

\subsection{The dependence of $\sigma_{rms}$ on luminosity and Eddington ratio}
\label{S:32}

\label{S:31}
\begin{figure}%[!h]
  \centering
  \includegraphics[width=\linewidth]{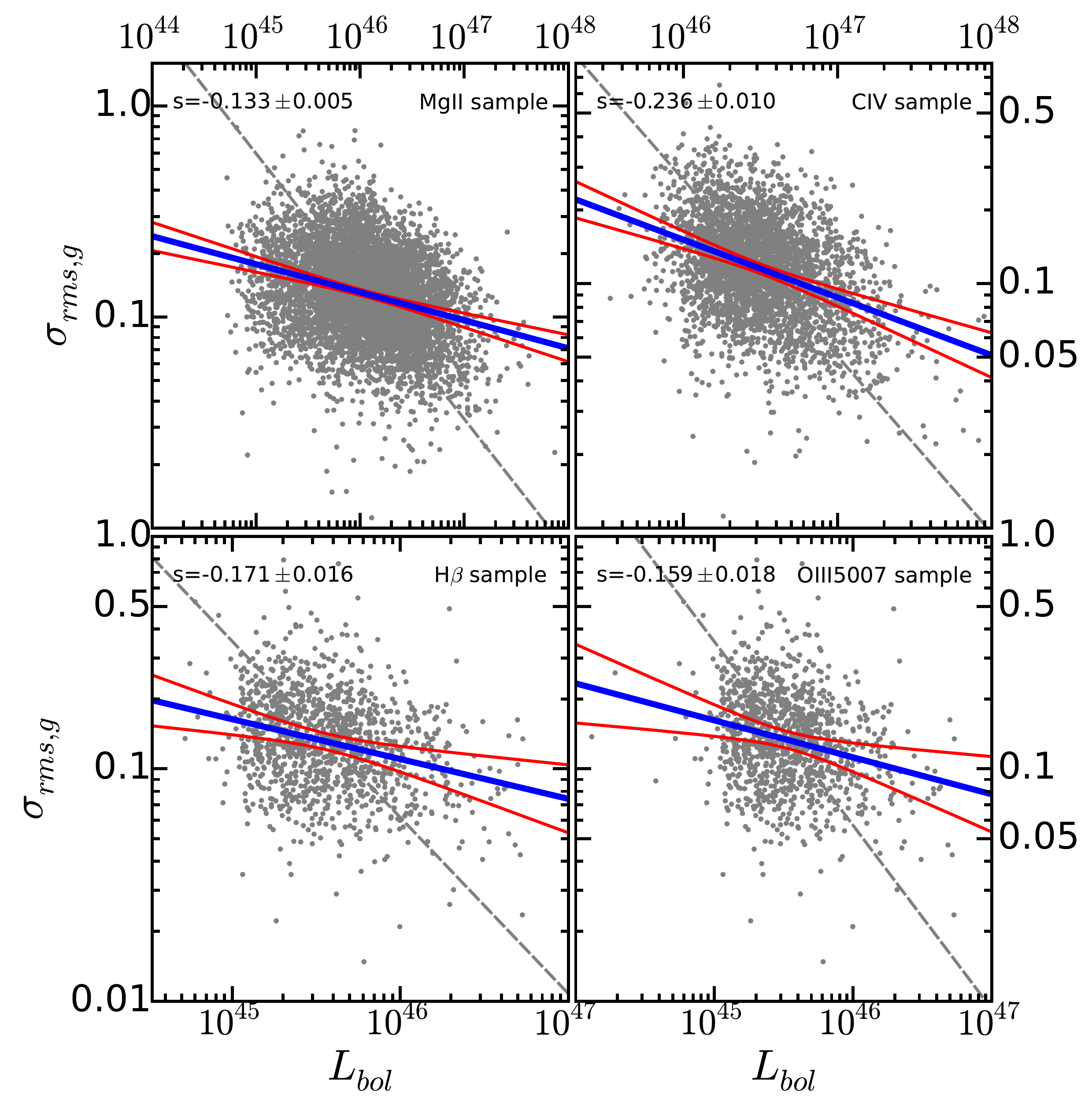}
  \caption{Correlations between $g$ band variability amplitude and bolometric luminosity for the four quasar samples. 
 Symbols and lines are the same as shown in Fig. \ref{Fig3_1}. In all samples, $\sigma_{rms}$ anti-correlates with $L_{bol}$. 
}
  \label{Fig3_1_2}
\end{figure}

\begin{table*}[h]
  \centering
  \caption{
  Correlations coefficients and linear regression slopes between
  between $\sigma_{rms}$ ($g$ band) and other parameters.}
  \begin{tabular}{c c c c c c}
    \hline
    \hline
       && broad MgII sample & CIV sample & H$\beta$ sample & [OIII]5007 sample\\
    \hline
       \multicolumn{6}{ c }{Pearson's Rank apparent correlation coefficients $r$, confidence levels $rcc$ and linear regression slopes $s$}\\
    \hline
       $L_{bol}$&r&-0.295(0.001)&-0.381(0.002)&-0.285(0.005)&-0.256(0.005)\\
       &rcc&$<$1e-16($<$1e-16,$<$1e-16)&$<$1e-16($<$1e-16,$<$1e-16)&$<$1e-16($<$1e-16,$<$1e-16)&$<$1e-16($<$1e-16,$<$1e-16)\\
       &s&-0.133$\pm$0.005&-0.236$\pm$0.010&-0.171$\pm$0.016&-0.159$\pm$0.018\\
    \hline
       $\frac{L_{bol}}{L_{Edd}}$&r&-0.329(0.008)&-0.205(0.009)&-0.293(0.017)&-0.284(0.016)\\
       &rcc&$<$1e-16($<$1e-16,$<$1e-16)&$<$1e-16($<$1e-16,$<$1e-16)&$<$1e-16($<$1e-16,$<$1e-16)&$<$1e-16($<$1e-16,$<$1e-16)\\
       &s&-0.173$\pm$0.006&-0.103$\pm$0.008&-0.129$\pm$0.012&-0.123$\pm$0.012\\
    \hline
       $M_{bh}$&r&-0.013(0.006)&-0.095(0.008)&0.080(0.018)&0.102(0.016)\\
       &rcc&0.14(0.06,0.28)&2e-8(1e-9,3e-7)&2e-3(3e-4,0.01)&3e-4(3e-5,2e-3)\\
       &s&-0.006$\pm$0.005&-0.043$\pm$0.008&0.034$\pm$0.012&0.043$\pm$0.012\\
    \hline
       $1+z$&r&-0.102(0.001)&-0.223(0.001)&-0.114(0.002)&-0.070(0.002)\\
       &rcc&1e-16($<$1e-16,1e-16)&$<$1e-16($<$1e-16,$<$1e-16)&3e-5(2e-5,4e-5)&9e-3(8e-3,0.01)\\
       &s&-0.233$\pm$0.028&-0.862$\pm$0.066&-0.556$\pm$0.139&-0.363$\pm$0.153\\
    \hline
       \multicolumn{6}{ c }{Partial correlation coefficients $r$ and confidence levels $rcc$}\\
    \hline
       $\frac{L_{bol}}{L_{Edd}}$($M_{bh}$, $1+z$)&r&-0.396(0.010)&-0.338(0.013)&-0.318(0.022)&-0.304(0.021)\\
       &rcc&$<$1e-16($<$1e-16,$<$1e-16)&$<$1e-16($<$1e-16,$<$1e-16)&$<$1e-16($<$1e-16,$<$1e-16)&$<$1e-16($<$1e-16,$<$1e-16)\\
    \hline
       $M_{bh}$($\frac{L_{bol}}{L_{Edd}}$, $1+z$)&r&-0.256(0.011)&-0.291(0.014)&-0.188(0.026)&-0.173(0.023)\\
       &rcc&$<$1e-16($<$1e-16,$<$1e-16)&$<$1e-16($<$1e-16,$<$1e-16)&2e-11(2e-14,6e-9)&2e-9(2e-11,2e-7)\\
    \hline
       $1+z$($\frac{L_{bol}}{L_{Edd}}$, $M_{bh}$)&r&0.175(0.010)&-0.072(0.005)&0.021(0.010)&0.028(0.009)\\
       &rcc&$<$1e-16($<$1e-16,$<$1e-16)&2e-5(4e-6,6e-5)&0.23(0.14,0.35)&0.17(0.11,0.26)\\
    \hline
       \multicolumn{6}{ c }{Multiple linear regression slopes (see Equation \ref{e:sigma})}\\
    \hline
       $\frac{L_{bol}}{L_{Edd}}$&a&-0.300$\pm$0.009&-0.252$\pm$0.012&-0.233$\pm$0.020&-0.221$\pm$0.021\\
       $M_{bh}$&b&-0.178$\pm$0.008&-0.198$\pm$0.011&-0.136$\pm$0.020&-0.125$\pm$0.021\\
       $1+z$&c&0.577$\pm$0.040&-0.282$\pm$0.068&0.118$\pm$0.157&0.160$\pm$0.167\\
    \hline
  \end{tabular}  
  \\
  \raggedright
Similar to Table \ref{Tab3_1}, but here the analyses are between $\sigma_{rms}$ and other parameters. 
$a$, $b$ and $c$ are slopes of the best-fit multiple linear regression in Equation \ref{e:sigma}. In this table, $\sigma_{rms}$ are in $g$ band. The $r$ and $i$ band results are similar, which are presented in Table \ref{Tab3_2_2} and Table \ref{Tab3_2_3} in Appendix.
  \label{Tab3_2}
\end{table*}

Following \S\ref{S:31}, we perform apparent correlation analyses between UV/optical variability amplitude $\sigma_{rms}$ and factors including 
bolometric luminosity, Eddington ratio $L_{bol}$/$L_{Edd}$, black hole mass $M_{bh}$, and $1+z$.
The results are shown in Table \ref{Tab3_2} (and Table \ref{Tab3_2_2} \& \ref{Tab3_2_3} in Appendix).
We plot the results of the four samples in Fig. \ref{Fig3_1_2}.
Clear negative correlations between UV/optical variability and luminosity are seen in all our samples, consistent with many literatures \citep[e.g.][]{berk2004ensemble, wilhite2008variability, 2010ApJ...716L..31A, Zuo2012, 2013A&A...560A.104M}. Besides, the variability also anti-correlates with the Eddington ratio
\citep[see also][]{berk2004ensemble, wilhite2008variability, 2010ApJ...716L..31A, Zuo2012, 2013A&A...560A.104M}. 
Apparent negative correlations between $\sigma_{rms}$ and redshift are also seen, primarily because that quasars at higher redshifts are generally more luminous.

 Again, to break the degeneracies between various parameters, partial correlation analyses are also performed (see Table \ref{Tab3_2}).
The partial correlation analyses show that $\sigma_{rms}$ anti-correlates with both  Eddington ratio and SMBH mass. 
The partial correlation between $\sigma_{rms}$ and redshift is primarily positive,  because AGN variation is known to be stronger at shorter rest frame wavelength (e.g., \citealt{
berk2004ensemble,Wilhite2005,Zuo2012,Sun2014,2011A&A...525A..37M,2018ApJ...868...58K}) and a given SDSS photometric band probes shorter rest frame wavelength for quasars at higher redshifts. 
The negative partial correlation between $g$ band $\sigma_{rms}$ and redshift in the CIV sample might be due to the fact that the strong Ly$\alpha$ line  (which is less variable than the continuum)
would be redshifted into $g$ band at redshift $>$ 2.3, making $g$ band variation weaker comparing with quasars at z $<$ 2.3. 
 Note that the  $\sigma_{rms}$ in this work measures the variability amplitude at certain timescale in the observed frame, therefore the time dilation effect exists here that for quasars at higher redshifts we are actually probe the variability at shorter timescale in the rest frame \citep[e.g.][]{Hawkins2010}.
Correcting the time dilation effect or the wavelength dependence of the variability is however  hard, as the exact relation between variability and timescale or wavelength is poorly constrained and may depend on other parameters. 
Fortunately, such effects would not affect the partial correlations between other parameters when the effect of redshift is controlled. 

We also perform multiple linear regression to quantify the relations between $\sigma_{rms}$ and the three physical parameters (see equation \ref{e:sigma} below, and the results in Table \ref{Tab3_2} (and Table \ref{Tab3_2_2} \& Table \ref{Tab3_2_3} in Appendix).
\begin{align}
 %\label{e4}
\sigma_{rms,exp} \sim (L_{bol}/L_{Edd})^a{M_{bh}}^b(1+z)^c
 \label{e:sigma}
 \end{align}
The results are consistent with partial correlation analyses.

\subsection{The intrinsic correlation between $EW$ and $\sigma_{rms}$}
\label{S:33}

\label{S:31}
\begin{figure}%[!h]
  \centering
  \includegraphics[width=\linewidth]{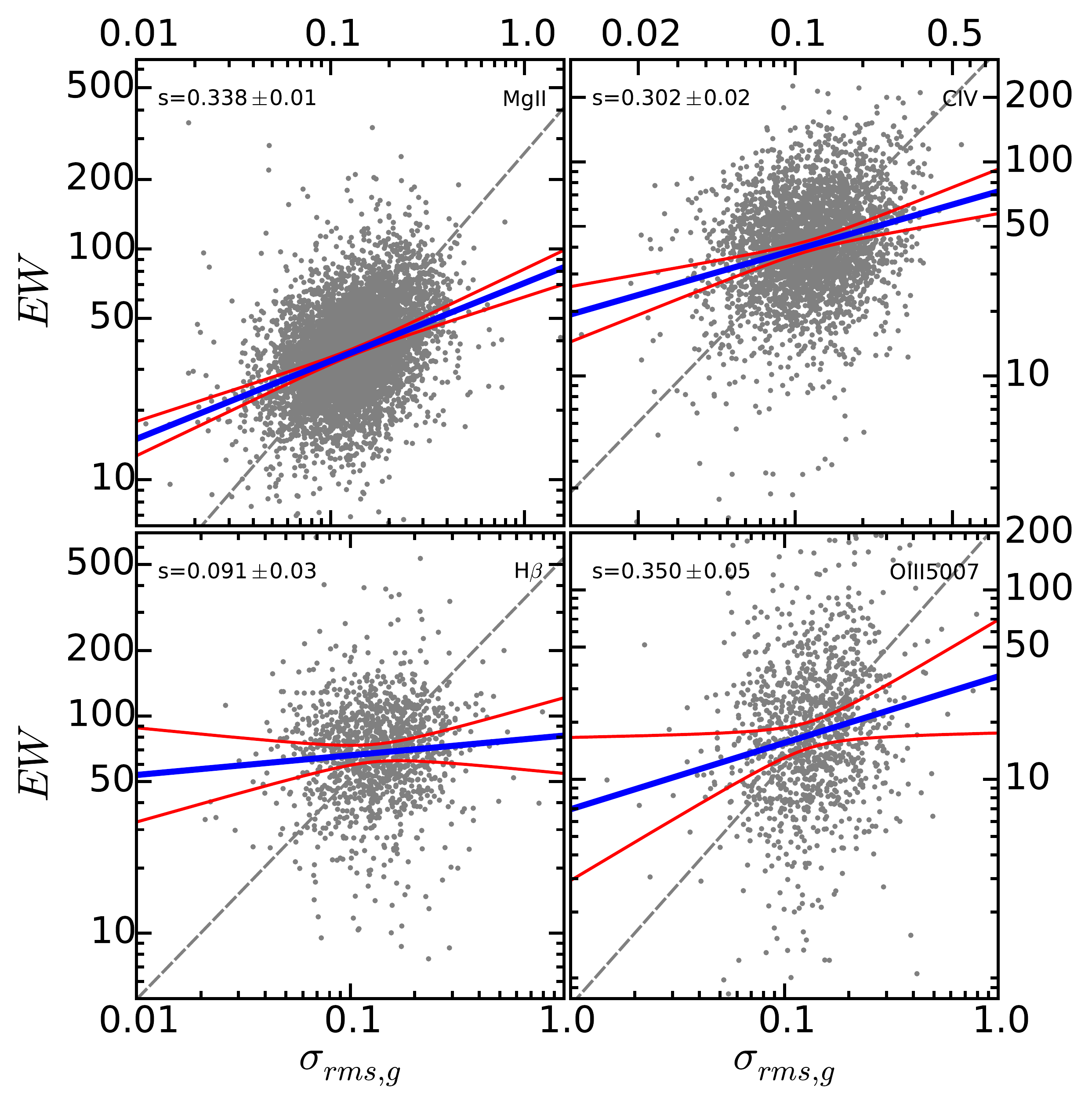}
  \caption{Correlations between line $EW$ and $g$ band variability amplitude. 
 Symbols and lines are the same as shown in Fig. \ref{Fig3_1}. 
}
  \label{Fig3_1_3}
\end{figure}

\begin{figure}%[!h]
  \centering
  \includegraphics[width=\linewidth]{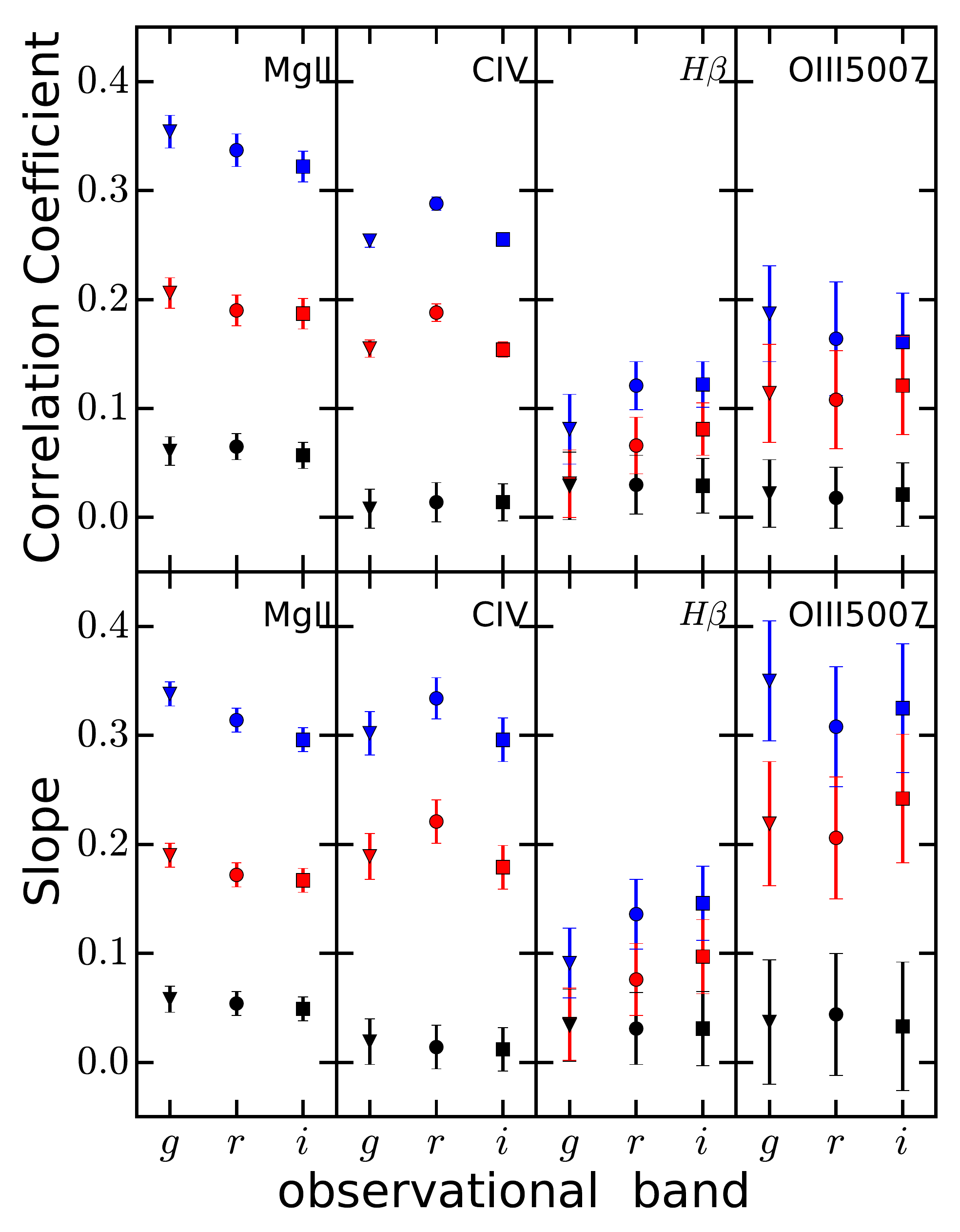}
  \caption{Correlation coefficient $r$ (upper panels) and best-fit linear regression slope $s$ (lower panels) between various line $EW$ and $\sigma_{rms}$.  
In each panel, the blue markers represent the apparent Pearson's correlation ($r$) and regression slope $s$ between $EW_{line}$ and $\sigma_{rms}$; the red ones represent the partial correlation coefficient $r$ (controlling Eddington ratio, black hole mass and redshift) and regression slope $s$ (Equation \ref{e3}); the black ones plotted the expected artificial correlations due to the uncertainties of the control variables if there is no intrinsic correlation between $EW_{line}$ and $\sigma_{rms}$. 
The inverted triangles, circles and squares stand for $g$, $r$ and $i$ observational band, respectively. 
}
  \label{Fig3_3}
\end{figure}

\begin{figure}%[!h]
  \centering
  \includegraphics[width=\linewidth]{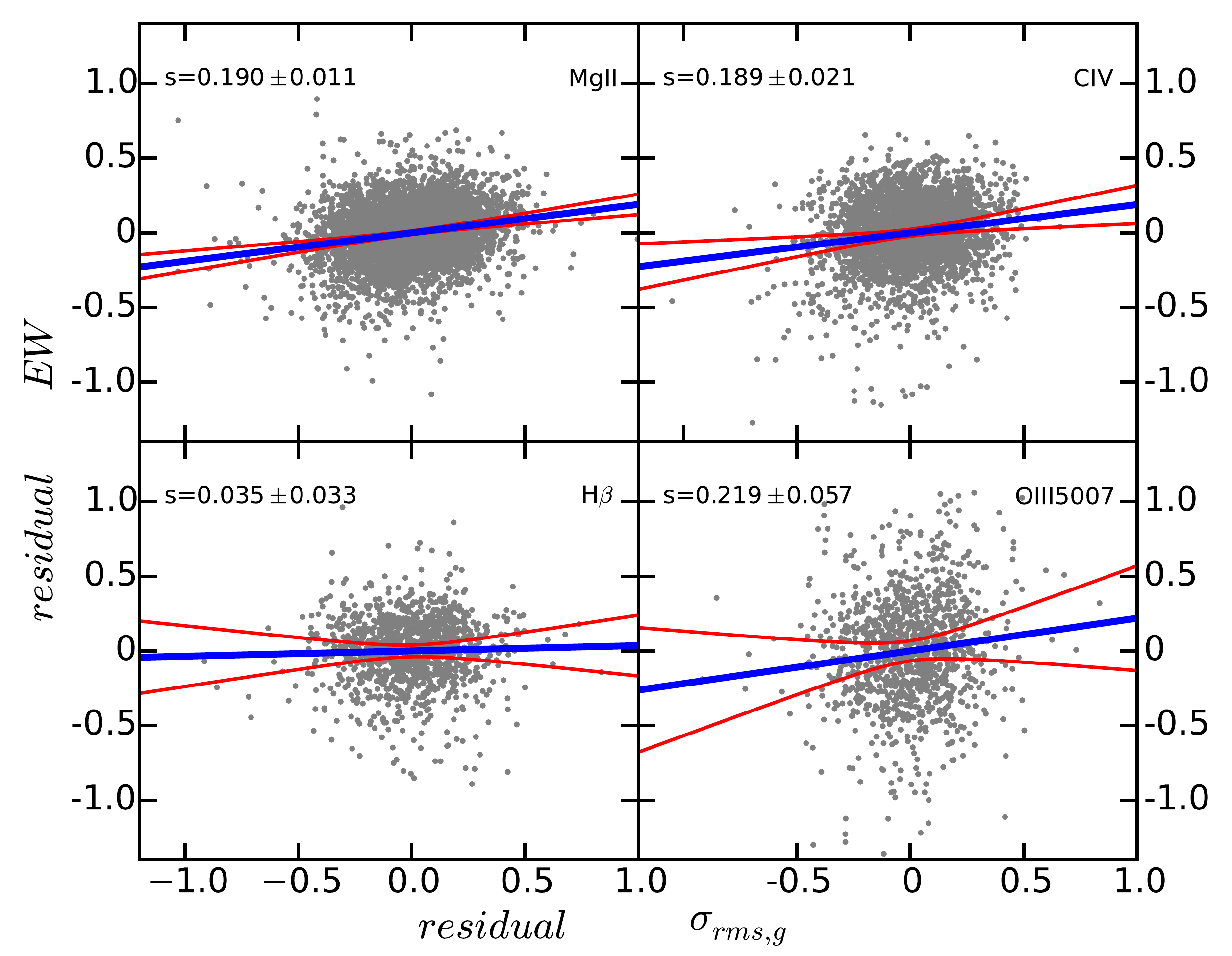}
  \caption{The residual $EW$ from equation \ref{e:EW} versus the residual $g$ band $\sigma_{rms}$ (results from $r$ and $i$ bands are rather similar) from equation \ref{e:sigma}, to demonstrate the intrinsic correlation between line $EW$ and $\sigma_{rms}$ after controlling the effects of Eddington ratio, black hole mass and redshift. Symbols and lines are the same as shown in Fig. \ref{Fig3_1}. The best-fit linear regression slopes, $s$, are given in the upper left corner in each panel, and the line names  in the upper right corner.
}
  \label{Fig3_2}
\end{figure}

\begin{table*}[!h]
  \centering
  \caption{Correlation coefficients and linear regression slopes between line $EW$ (named in the left-most column) and $\sigma_{rms}$ in various band (named in the right-most column)}
  \begin{tabular}{c c c c c c c c}
    \hline
    \hline
      Line Name & r & rcc & $L_{bol}/L_{Edd}$ (a) & $M_{BH}$ (b) & 1+z (c) & $\sigma_{rms}$ (s)& continuum band \\
    \hline
       \multicolumn{8}{ c }{Pearson's Rank apparent correlation coefficients $r$, confidence levels $rcc$ and linear regression slopes $s$ (between EW and $\sigma_{rms}$)}\\
    \hline
       &0.338(0.015)&$<10^{-16}$($<10^{-16}$,$<10^{-16}$)&&&&0.338$\pm$0.01&g\\
       broad MgII&0.320(0.015)&$<10^{-16}$($<10^{-16}$,$<10^{-16}$)&&&&0.314$\pm$0.01&r\\
       &0.308(0.014)&$<10^{-16}$($<10^{-16}$,$<10^{-16}$)&&&&0.296$\pm$0.01&i\\
    \hline
       &0.254(0.006)&$<10^{-16}$($<10^{-16}$,$<10^{-16}$)&&&&0.302$\pm$0.02&g\\
       CIV&0.288(0.006)&$<10^{-16}$($<10^{-16}$,$<10^{-16}$)&&&&0.334$\pm$0.02&r\\
       &0.255(0.005)&$<10^{-16}$($<10^{-16}$,$<10^{-16}$)&&&&0.296$\pm$0.02&i\\
    \hline
    	   &0.081(0.032)&$2\times 10^{-3}$($4\times 10^{-5}$,0.04)&&&&0.091$\pm$0.03&g\\
       H$\beta$&0.121(0.022)&$1\times 10^{-5}$($2\times 10^{-7}$,$2\times 10^{-4}$)&&&&0.136$\pm$0.03&r\\
       &0.122(0.021)&$1\times 10^{-5}$($2\times 10^{-7}$,$2\times 10^{-4}$)&&&&0.146$\pm$0.03&i\\
    \hline
       &0.187(0.052)&$1\times 10^{-10}$($2\times 10^{-16}$,$2\times 10^{-6}$)&&&&0.350$\pm$0.05&g\\
       $[$OIII$]$5007&0.164(0.045)&$2\times 10^{-8}$($6\times 10^{-13}$,$3\times 10^{-5}$)&&&&0.308$\pm$0.06&r\\
       &0.161(0.043)&$2\times 10^{-8}$($2\times 10^{-12}$,$3\times 10^{-5}$)&&&&0.325$\pm$0.06&i\\
    \hline
       \multicolumn{8}{ c }{Partial correlation coefficients ($r$ and $rcc$), and multiple linear regression slopes ($a,b,c,s$, see Equation \ref{e3})}\\
    \hline    
       &0.206(0.014)&$<10^{-16}$($<10^{-16}$,$<10^{-16}$)&-0.270$\pm$0.01&-0.119$\pm$0.01&0.556$\pm$0.04&0.190$\pm$0.01&g\\    
       broad MgII&0.190(0.014)&$<10^{-16}$($<10^{-16}$,$<10^{-16}$)&-0.275$\pm$0.01&-0.120$\pm$0.01&0.586$\pm$0.04&0.172$\pm$0.01&r\\    
       &0.187(0.014)&$<10^{-16}$($<10^{-16}$,$<10^{-16}$)&-0.280$\pm$0.01&-0.122$\pm$0.01&0.625$\pm$0.04&0.167$\pm$0.01&i\\    
    \hline
       &0.155(0.008)&$<10^{-16}$($<10^{-16}$,$<10^{-16}$)&-0.235$\pm$0.02&-0.228$\pm$0.01&0.522$\pm$0.08&0.189$\pm$0.02&g\\    
       CIV&0.188(0.008)&$<10^{-16}$($<10^{-16}$,$<10^{-16}$)&-0.224$\pm$0.02&-0.215$\pm$0.01&0.429$\pm$0.08&0.221$\pm$0.02&r\\    
       &0.154(0.007)&$<10^{-16}$($<10^{-16}$,$<10^{-16}$)&-0.236$\pm$0.02&-0.224$\pm$0.01&0.384$\pm$0.08&0.179$\pm$0.02&i\\    
    \hline
       &0.031(0.031)&0.14(0.02,0.5)&-0.106$\pm$0.02&-0.001$\pm$0.02&0.643$\pm$0.18&0.035$\pm$0.03&g\\    
       H$\beta$&0.066(0.026)&0.01($6\times 10^{-4}$,0.08)&-0.099$\pm$0.02&0.003$\pm$0.02&0.604$\pm$0.18&0.076$\pm$0.03&r\\    
       &0.081(0.024)&$2\times 10^{-3}$($1\times 10^{-4}$,0.02)&-0.101$\pm$0.02&0.000$\pm$0.02&0.617$\pm$0.18&0.097$\pm$0.03&i\\    
    \hline
       &0.114(0.045)&$6\times 10^{-5}$($4\times 10^{-8}$,0.01)&-0.261$\pm$0.04&-0.166$\pm$0.04&-0.049$\pm$0.32&0.219$\pm$0.06&g\\    
       $[$OIII$]$5007&0.108(0.043)&$1\times 10^{-4}$($2\times 10^{-7}$,0.01)&-0.271$\pm$0.04&-0.172$\pm$0.04&-0.153$\pm$0.32&0.206$\pm$0.06&r\\    
       &0.121(0.041)&$2\times 10^{-5}$($2\times 10^{-8}$,$4\times 10^{-3}$)&-0.280$\pm$0.04&-0.180$\pm$0.04&-0.097$\pm$0.32&0.242$\pm$0.06&i\\
    \hline
  \end{tabular}  
  \\
  \raggedright
This table lists the apparent Pearson's Rank correlation coefficients and confidence levels ($r$ and $rcc$) and the best-fit linear regression slopes $s$ between $EW$ and $\sigma_{rms}$ (when $a,b,c$ are not given).
When values of $a,b,c$ are listed,  $r$ and $rcc$ stand for the partial correlation coefficients ($r$ and $rcc$) between $EW$ and $\sigma_{rms}$ after controlling the effects of Eddington ratio, black hole mass and redshift, and $a,b,c$ and $s$ the best-fit multiple linear regression slopes in Equation \ref{e3}.
  \label{Tab3_3}
\end{table*}

Since both line $EW$s and $\sigma_{rms}$ similarly anti-correlate with luminosity and $L_{bol}$/$L_{Edd}$, it is not surprising that the two quantities show apparent positive correlations (see Table \ref{Tab3_3} and Fig. \ref{Fig3_1_3}). Partial correlation analyses are thus essential to reveal the intrinsic correlation between the two quantities. 
 In Table \ref{Tab3_3} we also present the partial correlation coefficients between line $EW$ and $\sigma_{rms}$, by controlling the effects of $L_{bol}/L_{Edd}$, $M_{bh}$ and redshift. 
Again, as $L_{bol}/L_{Edd}$ is simply the ratio of $L_{bol}$ and $M_{bh}$, the effect of $L_{bol}$ is also simultaneously controlled during the analyses. Replacing $L_{bol}/L_{Edd}$ with $L_{bol}$ during the analyses does not alter the results. 

Partial correlation analyses reveal strong intrinsic correlations between line $EW$ of broad MgII, CIV and $\sigma_{rms}$ ($g,r,i$), though with coefficients considerably smaller than the apparent correlations (see Fig. \ref{Fig3_3}). 
Such intrinsic correlations indicate that at given Eddington ratio, black hole mass and redshift, quasars with stronger variabilities in UV/optical have stronger broad MgII and CIV emission lines. The intrinsic correlation coefficient $r$ between [OIII]5007 $EW$ and $\sigma_{rms}$ is smaller but still statistically significant, 
and that between broad H$\beta$ and $\sigma_{rms}$ is the weakest among the four lines. 

Meanwhile, we also perform multiple linear regression analyses to quantify the relations between line $EW$ and physical parameters including Eddington ratio, black hole mass, redshift, and UV/optical variability amplitude $\sigma_{rms}$, 
 \begin{align}
 %\label{e3}
EW \sim (L_{bol}/L_{Edd})^a{M_{bh}}^b(1+z)^c{\sigma_{rms}}^s
 \label{e3}
 \end{align}
 The results of best-fit parameters are displayed in Table \ref{Tab3_3}, showing intrinsic correlation patterns between line $EW$ and $\sigma_{rms}$ consistent with those from partial correlation analyses.

To directly illustrate the intrinsic correlation between line $EW$ and $\sigma_{rms}$,  we derive the residual line $EW$ with respect to the best-fit equation \ref{e:EW}
and the residual $\sigma_{rms}$ with respect to the best-fit equation \ref{e:sigma}, and plot them in Fig. \ref{Fig3_2}. 
The linear regression slopes in the figure are similar with the ones between $EW$ and $\sigma_{rms}$ from multiple linear regression in Table \ref{Tab3_3}. 
We note large scatter is clearly visible in the plot (see also Fig. \ref{Fig3_1_3}). We exact the outliers (5\% above and 5\% below, with the largest perpendicular distance to the regression line) from the plot,
and find the outliers show statistically indistinguishable distributions of $L_{bol}$, $M_{bh}$ and corresponding line width, compared with the whole sample. They also generally have normal SDSS spectra, except for that some outliers lie very close to the redshift limits of each sample (likely because SDSS spectral quality is worse near the red/blue ends). Excluding sources close to the redshift limits would not alter the results in this work. 

 It's well known that the measurements of black hole mass and bolometric luminosity of quasars suffer from considerable uncertainties.
The large uncertainties in the control variables may lead to artificial partial correlations between two quantities which both correlate with the control variables. 
Following \cite{2018ApJ...868...58K}, we perform simulations to examine possible artificial partial correlation due to the uncertainties of $L_{bol}/L_{Edd}$ and $M_{bh}$. 
Utilizing the observed $L_{bol}/L_{Edd}$, $M_{bh}$ and redshift for each quasar, we calculate the expected line $EW$ and $\sigma_{rms}$ based on the best-fit 
equation \ref{e:EW} \& \ref{e:sigma} respectively. Random Gaussian fluctuations are then added to the expected values to reproduce the observed scatters in equation \ref{e:EW} \& \ref{e:sigma}. 
The simulated line $EW$ and $\sigma_{rms}$ we produced have no intrinsic correlation between each other. However, after we randomly fluctuate the values of $L_{bol}/L_{Edd}$, $M_{bh}$ to mimic their measurement uncertainties, artificial partial correlation between line $EW$ and $\sigma_{rms}$ could emerge. 
For $L_{bol}$, we adopt a 0.08 dex uncertainty (20\%, to take account of the uncertainty in bolometric correction, \citealt{2006ApJS..166..470R}), and add it quadratically to the direct measurement error from \citep{shen2011catalog}. For mass measurement, both a conservative 0.4 dex calibration uncertainty \citep{shen2011catalog} and the direct measurement error from \cite{shen2011catalog} are included. No fluctuation is added to redshift as it has considerably small uncertainty. Partial correlation analyses using the simulated samples do show positive partial correlations between line $EW$ and $\sigma_{rms}$, 
but too weak to explain the observed correlations for MgII, CIV and [OIII]5007 (see Fig. \ref{Fig3_3}).

We finally note that in this work, the line $EW$, bolometric luminosity and SMBH mass for each quasar (from \citealt{shen2011catalog}) are measured based on single-epoch SDSS spectra obtained at certain spectral MJD (sMJD), while $\sigma_{rms}$ is measured over a period of $\sim$ 10 years. We show below such fact does not bias the results in this work.
Comparing sMJD with photometric observations for our sample, we find on average $\sim$ 80\% of the photometric data points were obtained after sMJD. 
We further compare the synthetic photometry measured from the spectra with the photometric data points,  and find $\sim$ 15\% of our quasars have the synthetic photometry fainter than the minimum brightness in the corresponding photometric light curve (but contrarily 4.7\% of quasars have the synthetic photometry brighter than the maximum brightness in the light curve). Those are likely due to the fiber-drop issue \citep[e.g.][]{Guo2020}. Excluding those sources however does not alter the results of this work. Other than those sources with fiber-dropping, we do not find systematic offset between the photometric and the synthetic photometry, i.e., the SDSS spectra could represent the properties of the quasars at a random epoch.
Furthermore, around half of our quasars have repeated SDSS spectroscopy. Our results remain unchanged if we utilize the spectra other than those used by  \cite{shen2011catalog} and measure the corresponding line $EW$, bolometric luminosity and SMBH mass following an approach similar to Shen et al. (Ren et al. in preparation).

\section{Discussion}
\label{S:discussion}

The intrinsic correlations we have revealed between the strong emission line (MgII, CIV and [OIII]5007, but not H$\beta$) $EW$ of quasars and UV/optical variability amplitude
 indicate that more variable quasars have stronger emission lines\footnote{ Since during the partial correlation analyses, the effects of bolometric luminosity (simply derived from continuum luminosity) and redshift have been removed, a partial correlation between line $EW$ and variability also means a partial correlation between line flux (or luminosity) and variability. This is confirmed through replacing line $EW$ with flux (or luminosity) during the analyses.}.
 Note that \cite{Rumbaugh2018} found that extreme variable quasars (those with a maximum change in g-band magnitude of more than 1 mag) tend to have stronger emission lines (MgII, CIV and [OIII]5007) compared with a control sample with matched luminosity and redshift (see also Ren et al. in prep.), nicely consistent with our findings. 

The correlations show that the line production and disc turbulence are physically connected.
Below we first propose two interesting mechanisms behind such intrinsic correlations: 1) stronger disc turbulence yields bluer/harder quasar SED, thus stronger emission lines; 2) disc magnetic turbulences launch outflowing wind which could elevate the covering factor of BLR and NLR clouds. We finally briefly discuss the puzzling different behavior of H$\beta$ line (compared with MgII, CIV and [OIII]5007). 

Theoretically, disc thermal fluctuating models \citep{Dexter2011, Cai2016, Cai2018, Cai2019, 2020ApJ...892...63C} associate the multi-wavelength variability to magnetic turbulence in the accretion disc. Note such fluctuation models indeed predict bluer averaged EUV SEDs than the standard thin disk model without temperature fluctuation, and the stronger turbulence the bluer the mean SED \citep[see Fig. 4 of][]{Cai2016}. 
This is qualitatively consistent with the discovery presented in this work that quasars with stronger UV/optical variability have stronger emission lines, 
albeit it is yet to be observationally confirmed whether quasars which are more variable do have blue extreme UV SED (Cai et al. in prep).  
As reproducing the ionizing SED of AGNs is never straightforward, we would defer a quantitative comparison with predictions of disc fluctuation model and our results to a future work. 
Meanwhile we have previously found a positive correlation between the UV/optical variability and the X-ray loudness \citep{2018ApJ...868...58K}, showing the X-ray corona heating in AGNs could be also closely associated with magnetic turbulence, and suggesting more variable quasars do have relatively harder SED which could produce stronger emission lines. 

Alternatively, stronger disc magnetic turbulences might be able to launch disc winds with larger covering factor, thus yielding stronger emission lines. 
Please refer to \S\ref{introduction} for references of theoretical models of disc winds. 
While it is extremely challenging to theoretically depict the role of magnetic turbulence in wind launch, 
this work brings up an interesting potential probe of it: comparing the observational properties of AGNs with stronger disc turbulences with those quieter ones to probe the sequence of the disc turbulence. For instance, in additional to emission line $EW$, one may investigate the connection between disc turbulence and emission line profile (Ren et al. in prep).
 
However, it is yet difficult to distinguish the two scenarios we proposed above, i.e., bluer/harder SED or larger covering factor of emission line clouds. 
We note that compared with broad MgII and CIV lines, [OIII]5007 shows weaker intrinsic correlation (the partial correlation coefficient $r$) with $\sigma_{rms}$ (see Table \ref{Tab3_3} and Fig. \ref{Fig3_3}). This is likely because compared with BLR, the covering factor of NLR could be affected by additional factors such as the torus and the ISM environment thus showing significantly larger scatter (see also Fig. \ref{Fig3_2}). 
 Further note that OIII line comes from the narrow line region, i.e., averaging variability over thousand years, while $\sigma_{rms}$ in this paper is measured with a decade timescale.
This fact could also play a role in producing the large scatter and reducing the correlation coefficient $r$ between [OIII]5007 $EW$ and $\sigma_{rms}$.
Notably, comparing with MgII and CIV, [OIII]5007 exhibits similar linear regression slope with $\sigma_{rms}$ (Fig. \ref{Fig3_3}). Such fact tends to favor the bluer/harder SED scenario, as both BLR and NLR are illuminated and expected to be ionized by the same central radiation. However, though NLR has much large physical scale, the turbulence-driven disc wind scenario can not be ruled out if such wind could eventually reach the NLR \citep[e.g.][]{Proga2008}. For instance, \cite{Du2014} reported a strong correlation between BLR and NLR metallicities in AGNs, suggesting outflows from BLRs could carry metal rich gas to NLRs. 
 
\begin{figure}%[!h]
  \centering
  \includegraphics[width=\linewidth]{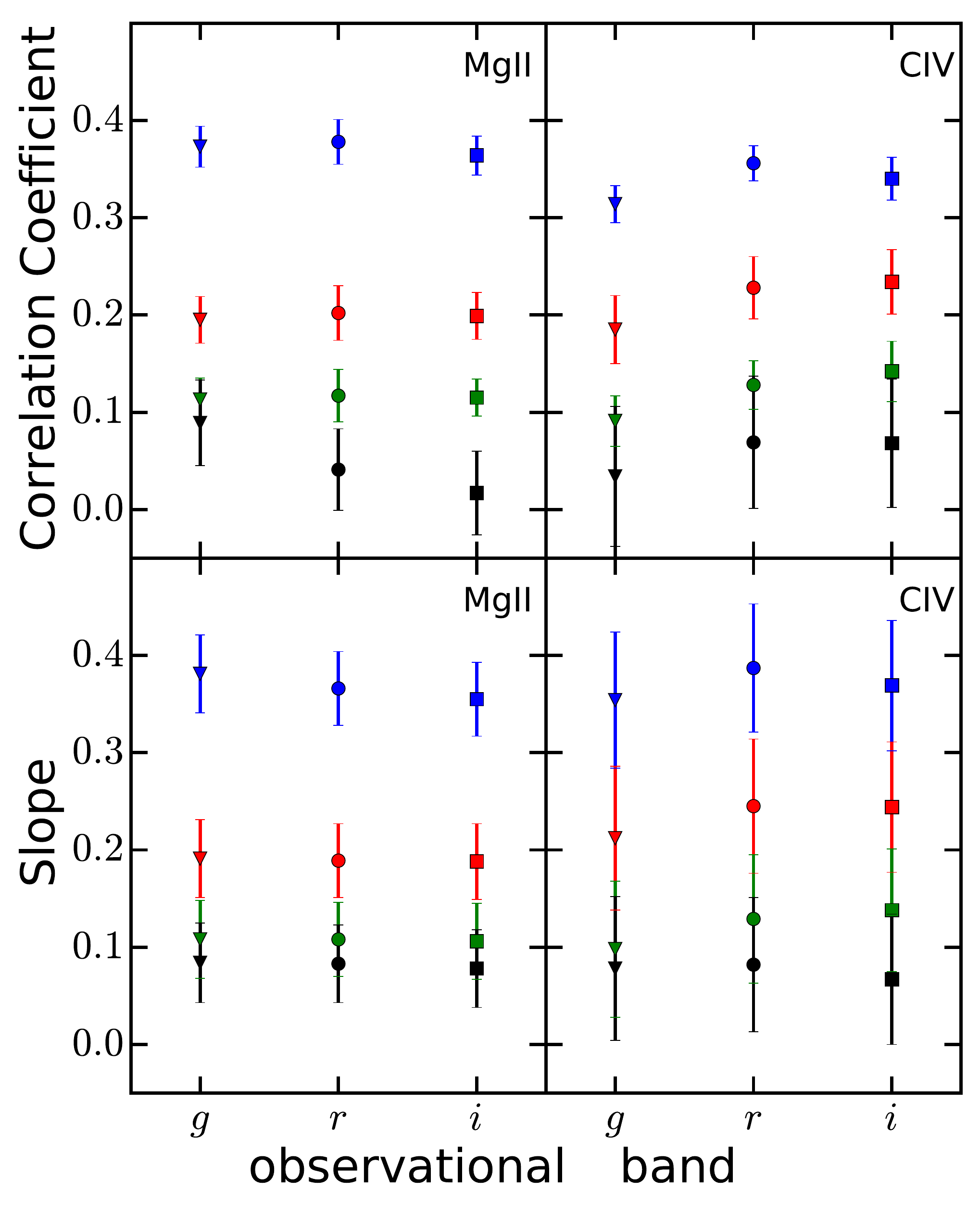}
  \caption{
  Similar to Fig. \ref{Fig3_3}, but for small subsamples (of MgII and CIV) with X-ray detections.   
In each panel, the blue markers represent the apparent Pearson's correlation ($r$) and regression slope $s$ between $EW_{line}$ and $\sigma_{rms}$; the red ones represent the partial correlation coefficient $r$ (controlling Eddington ratio, black hole mass and redshift) and regression slope $s$ (Equation \ref{e3}); the green ones represent partial correlations after further controlling X-ray loudness; and
 the black ones plotted the expected artificial correlations due to the uncertainties of the control variables if there is no intrinsic correlation between $EW_{line}$ and $\sigma_{rms}$. 
}
  \label{Fig4_2}
\end{figure}

Due to the lack of extreme UV coverage, it is hard to constrain the ionizing SED of quasars.  As an experiment, below we adopt X-ray loudness as an approximate proxy of broadband SED to investigate whether
harder SED\footnote{Note \cite{Wu2009} did report a positive correlation between CIV line $EW$ and the relative X-ray to UV brightness.}
 could fully account for the observed intrinsic correlation between line $EW$ and $\sigma_{rms}$.
Note a caveat of this approach is that 
quasars with the same X-ray loudness do not necessarily have the same EUV SED.  
To derive the X-ray loudness of our quasars, following \cite{2018ApJ...868...58K}
we cross-match our MgII and CIV samples with the Stripe 82X X-ray source catalog \citep{ananna2017agn}. The source number of X-ray matched MgII and CIV samples are 572 and 236, respectively. The significant reduction of the sample sizes is due to the limited coverage of Stripe 82X (31.3 vs 290 deg$^2$), and the X-ray detection completeness of the parent samples within the Stripe 82X footprint is considerably high (66\% - 80\%)\footnote{
 Following \cite{2018ApJ...868...58K} we estimate the effect of X-ray sample incompleteness and conclude that the incompleteness does not affect the results presented below.
}. 
The [OIII]5007 and H$\beta$ samples are excluded because the final X-ray matched samples are too small. 
We then calculate the X-ray loudness ($L_{0.5 - 10 keV}/L_{bol}$) for the X-ray detected subsamples. 
In Fig \ref{Fig4_2}, we plot the correlation coefficients $r$ between line $EW$ and $\sigma_{rms}$ for the X-ray detected subsamples. Similar to what we have seen in the parent samples, the partial correlations between $EW$ and $\sigma_{rms}$ are evident for the X-ray subsamples, after controlling for the effect of Eddington ratio, black hole mass and redshift, though considerably weaker than the apparent correlations.
Further controlling the effect of  X-ray loudness ulteriorly reduces the correlation coefficients (Fig \ref{Fig4_2}),  indicating that the first mechanism (more variable quasars have harder SED) may have played a significant role. 

We then examine whether the residual partial correlations between $EW$ and $\sigma_{rms}$ could be artificial due to uncertainties in the control variables. 
Again, we adopt a 0.08 dex calibration uncertainty for $L_{bol}$ and 0.4 dex calibration uncertainty for $M_{bh}$ in additional to their statistical observational uncertainties from \cite{shen2011catalog}.
For X-ray luminosity used in the calculation of X-ray loudness, we employ 0.08 dex as the mean observational uncertainty (since not all X-ray sources have flux errors in the Stripe 82X catalog). 
\cite{Middei2017} provided the long term X-ray variation of SDSS quasars, and the structure function at $\sim$ 10 years is $\sim$ 0.3 dex.  
We further include a 0.3/$\sqrt{2}$ dex to represent the random long term variability of X-ray fluxes in SDSS quasars.
The simulated artificial partial correlation coefficients (black data points in Fig \ref{Fig4_2}) are smaller than though statistically comparable to the residual coefficients (green data points in Fig \ref{Fig4_2}). This suggests the variation of SED alone might be insufficient to fully account for the intrinsic correlation between line $EW$ and $\sigma_{rms}$ we reported in this work.
The second mechanism (stronger disc turbulence launches emission line clouds with larger sky coverage) may also be involved.

Finally, it is worth noting that H$\beta$ $EW$ shows no (or at most marginal) partial correlation with $\sigma_{rms}$, making it distinct from other lines (see Fig. \ref{Fig3_3}).
Note H$\beta$ also shows very weak or no Baldwin effect \citep[see Table \ref{Tab3_1}, and ][]{1999AJ....118.2658S, 2002ApJ...581..912D, 2007ApJ...671..104L, 2017A&A...603A..49R}.
Considering the Baldwin effect is prominent for Ly$\alpha$ line while it is missing for Balmer lines, \cite{2002ApJ...581..912D} proposed that the different behavior of Ly$\alpha$ and Balmer lines 
could be related to the complicated physical processes of Ly$\alpha$ and H$\beta$ line emission \citep[e.g.][]{Fabian1995, 2020MNRAS.494.1611N}. The same mechanism may also account for 
the different behavior of H$\beta$ in the $EW$ $\sim$ $\sigma_{rms}$ relation comparing with other lines. 

\section{Conclusions}

In this work, we investigate the correlation between emission line (broad MgII, CIV, [OIII]5007 and broad H$\beta$) $EW$ and UV/optical variability amplitude $\sigma_{rms}$ for SDSS Stripe 82 quasars.  We show the two quantities show clear apparent correlations. Meanwhile both quantities show apparent anti-correlations with bolometric luminosity and Eddington ratio. 

We perform partial correlation analyses and reveal intrinsic correlations between line $EW$ (of MgII, CIV and [OIII]5007) and $\sigma_{rms}$, after controlling for the effect of luminosity, Eddington ratio, black hole mass and redshift. 
Interestingly, broad H$\beta$, of which the Baldwin effect is  known to be absent, doe not show clear intrinsic correlation between $EW$ and $\sigma_{rms}$ either.

The intrinsic correlations between line $EW$ (of MgII, CIV and [OIII]5007) and UV/optical variability amplitude suggest their underlying processes, i.e., line production and disc turbulence, are physically connected. We propose two possible mechanisms, both may be involved, for such connection: 1) more variable quasars tend to have bluer/harder SED; 2) more variable quasars can launch emission line clouds with larger covering factor.

\section*{Acknowledgement}
The work is supported by National Natural Science Foundation of China (grants No. 11421303, 11890693, 12033006 $\&$ 11873045) and CAS Frontier Science Key Research Program (QYZDJ-SSW-SLH006).

\bibliography{qv_170212_arxiv.bbl}

\appendix

\begin{table*}[h]
  \centering
  \caption{
  Correlations coefficients and linear regression slopes between
  between $\sigma_{rms}$ (in $r$ band) and other parameters.}
  \begin{tabular}{c c c c c c}
    \hline
    \hline
       && broad MgII & CIV & H$\beta$ & [OIII]5007\\
    \hline
       \multicolumn{6}{ c }{Pearson's Rank apparent correlation coefficients $r$, confidence levels $rcc$ and linear regression slopes $s$}\\    
    \hline
       $L_{bol}$&r&-0.342(0.002)&-0.363(0.002)&-0.199(0.006)&-0.164(0.006)\\
       &rcc&$<$1e-16($<$1e-16,$<$1e-16)&$<$1e-16($<$1e-16,$<$1e-16)&1e-12(2e-13,5e-12)&1e-8(4e-9,4e-8)\\
       &s&-0.158$\pm$0.005&-0.230$\pm$0.010&-0.118$\pm$0.017&-0.101$\pm$0.018\\
    \hline
       $\frac{L_{bol}}{L_{Edd}}$&r&-0.322(0.008)&-0.169(0.010)&-0.265(0.016)&-0.255(0.016)\\
       &rcc&$<$1e-16($<$1e-16,$<$1e-16)&$<$1e-16($<$1e-16,2e-16)&$<$1e-16($<$1e-16,$<$1e-16)&$<$1e-16($<$1e-16,2e-16)\\
       &s&-0.173$\pm$0.006&-0.087$\pm$0.009&-0.116$\pm$0.012&-0.110$\pm$0.012\\
    \hline
       $M_{bh}$&r&-0.065(0.007)&-0.114(0.008)&0.116(0.017)&0.137(0.015)\\
       &rcc&7e-8(3e-9,1e-6)&2e-11(9e-13,5e-10)&2e-5(1e-6,2e-4)&2e-6(1e-7,2e-5)\\
       &s&-0.029$\pm$0.005&-0.053$\pm$0.008&0.049$\pm$0.012&0.058$\pm$0.012\\
    \hline
       $1+z$&r&-0.162(0.002)&-0.119(0.001)&-0.003(0.004)&0.050(0.004)\\
       &rcc&$<$1e-16($<$1e-16,$<$1e-16)&3e-12(2e-12,5e-12)&0.46(0.40,0.49)&0.05(0.03,0.06)\\
       &s&-0.380$\pm$0.029&-0.471$\pm$0.068&-0.013$\pm$0.139&0.255$\pm$0.153\\
    \hline
       \multicolumn{6}{ c }{Partial correlation coefficients $r$ and confidence levels $rcc$}\\    
    \hline
       $\frac{L_{bol}}{L_{Edd}}$($M_{bh}$, $1+z$)&r&-0.391(0.010)&-0.344(0.014)&-0.283(0.019)&-0.261(0.021)\\
       &rcc&$<$1e-16($<$1e-16,$<$1e-16)&$<$1e-16($<$1e-16,$<$1e-16)&$<$1e-16($<$1e-16,$<$1e-16)&$<$1e-16($<$1e-16,1e-16)\\
    \hline
       $M_{bh}$($\frac{L_{bol}}{L_{Edd}}$, $1+z$)&r&-0.268(0.011)&-0.322(0.014)&-0.158(0.022)&-0.138(0.022)\\
       &rcc&$<$1e-16($<$1e-16,$<$1e-16)&$<$1e-16($<$1e-16,$<$1e-16)&1e-8(1e-10,9e-7)&2e-6(3e-8,5e-5)\\
    \hline
       $1+z$($\frac{L_{bol}}{L_{Edd}}$, $M_{bh}$)&r&0.140(0.010)&0.043(0.005)&0.101(0.009)&0.119(0.008)\\
       &rcc&$<$1e-16($<$1e-16,$<$1e-16)&6e-3(3e-3,0.01)&2e-4(6e-5,6e-4)&3e-5(9e-6,9e-5)\\
    \hline
       \multicolumn{6}{ c }{Multiple linear regression slopes (see Equation \ref{e:sigma}}\\
    \hline
       $\frac{L_{bol}}{L_{Edd}}$&a&-0.301$\pm$0.009&-0.267$\pm$0.013&-0.207$\pm$0.020&-0.189$\pm$0.021\\
       $M_{bh}$&b&-0.190$\pm$0.008&-0.229$\pm$0.012&-0.115$\pm$0.020&-0.099$\pm$0.021\\
       $1+z$&c&0.466$\pm$0.041&0.176$\pm$0.071&0.565$\pm$0.159&0.675$\pm$0.168\\
    \hline
  \end{tabular}  
  \\
  \raggedright
Similar to Table \ref{Tab3_2}, but with $\sigma_{rms}$ in $r$ band. The results are similar to Table \ref{Tab3_2}.
  \label{Tab3_2_2}
\end{table*}

\begin{table*}[h]
  \centering
  \caption{
  Correlations coefficients and linear regression slopes between
  between $\sigma_{rms}$ (in $i$ band) and other parameters.}
  \begin{tabular}{c c c c c c}
    \hline
    \hline
       && broad MgII & CIV & H$\beta$ & [OIII]5007\\
    \hline
       \multicolumn{6}{ c }{Pearson's Rank apparent correlation coefficients $r$, confidence levels $rcc$ and linear regression slopes $s$}\\    
    \hline
       $L_{bol}$&r&-0.380(0.008)&-0.332(0.003)&-0.136(0.013)&-0.113(0.010)\\
       &rcc&$<$1e-16($<$1e-16,$<$1e-16)&$<$1e-16($<$1e-16,$<$1e-16)&9e-7(8e-8,8e-6)&7e-5(2e-5,2e-4)\\
       &s&-0.178$\pm$0.005&-0.210$\pm$0.010&-0.076$\pm$0.016&-0.065$\pm$0.017\\
    \hline
       $\frac{L_{bol}}{L_{Edd}}$&r&-0.313(0.010)&-0.151(0.011)&-0.202(0.028)&-0.197(0.025)\\
       &rcc&$<$1e-16($<$1e-16,$<$1e-16)&$<$1e-16($<$1e-16,$<$1e-16)&5e-13(1e-16,4e-10)&1e-11(2e-14,3e-9)\\
       &s&-0.171$\pm$0.006&-0.078$\pm$0.009&-0.082$\pm$0.011&-0.080$\pm$0.012\\
    \hline
       $M_{bh}$&r&-0.108(0.007)&-0.107(0.010)&0.099(0.022)&0.116(0.022)\\
       &rcc&$<$1e-16($<$1e-16,1e-16)&3e-10(7e-12,1e-8)&3e-4(1e-5,3e-3)&5e-5(2e-6,8e-4)\\
       &s&-0.049$\pm$0.006&-0.050$\pm$0.008&0.039$\pm$0.011&0.046$\pm$0.012\\
    \hline
       $1+z$&r&-0.232(0.006)&-0.045(0.001)&-0.007(0.003)&0.023(0.005)\\
       &rcc&$<$1e-16($<$1e-16,$<$1e-16)&5e-3(4e-3,6e-3)&0.40(0.36,0.44)&0.22(0.17,0.27)\\
       &s&-0.554$\pm$0.029&-0.178$\pm$0.069&-0.033$\pm$0.130&0.110$\pm$0.143\\
    \hline
       \multicolumn{6}{ c }{Partial correlation coefficients $r$ and confidence levels $rcc$}\\    
    \hline
       $\frac{L_{bol}}{L_{Edd}}$($M_{bh}$, $1+z$)&r&-0.366(0.011)&-0.339(0.015)&-0.196(0.027)&-0.180(0.026)\\
       &rcc&$<$1e-16($<$1e-16,$<$1e-16)&$<$1e-16($<$1e-16,$<$1e-16)&2e-12(1e-15,1e-9)&5e-10(1e-12,1e-7)\\
    \hline
       $M_{bh}$($\frac{L_{bol}}{L_{Edd}}$, $1+z$)&r&-0.252(0.011)&-0.323(0.016)&-0.095(0.022)&-0.078(0.022)\\
       &rcc&$<$1e-16($<$1e-16,$<$1e-16)&$<$1e-16($<$1e-16,$<$1e-16)&4e-4(2e-5,5e-3)&4e-3(4e-4,0.03)\\
    \hline
       $1+z$($\frac{L_{bol}}{L_{Edd}}$, $M_{bh}$)&r&0.073(0.011)&0.115(0.006)&0.058(0.008)&0.064(0.008)\\
       &rcc&2e-9(5e-12,2e-7)&2e-11(1e-12,2e-10)&0.02(0.01,0.04)&0.02(8e-3,0.03)\\
    \hline
       \multicolumn{6}{ c }{Multiple linear regression slopes (see Equation \ref{e:sigma}}\\
    \hline
       $\frac{L_{bol}}{L_{Edd}}$&a&-0.283$\pm$0.009&-0.264$\pm$0.013&-0.134$\pm$0.019&-0.122$\pm$0.020\\
       $M_{bh}$&b&-0.180$\pm$0.008&-0.232$\pm$0.012&-0.065$\pm$0.020&-0.054$\pm$0.020\\
       $1+z$&c&0.244$\pm$0.041&0.472$\pm$0.071&0.309$\pm$0.152&0.344$\pm$0.161\\
    \hline
  \end{tabular}  
  \\
  \raggedright
Similar to Table \ref{Tab3_2}, but with $\sigma_{rms}$ in $i$ band. The results are similar to Table \ref{Tab3_2}.
  \label{Tab3_2_3}
\end{table*}

\end{document}